\documentclass[%
 reprint,
 amsmath,amssymb,
 prb,
]{revtex4-1}

\usepackage{graphicx}
\usepackage{dcolumn}
\usepackage{bm}
\usepackage{float} 
\begin{document}

\preprint{APS/123-QED}

\title{Comment on ``Combined experimental and computational study of the recrystallization process induced by electronic interactions\\ of swift heavy ions with silicon carbide crystals''}

\author{A. Benyagoub}
 
\affiliation{%
 Centre de Recherche sur les Ions, les Mat\'eriaux et la Photonique (CIMAP, ex CIRIL-GANIL), \\ CEA-CNRS-ENSICAEN-Universit\'e de Caen
 Bd Henri Becquerel, BP 5133, F-14070 Caen Cedex 5, France.}%

\date{February 13, 2014}

\begin{abstract}
\medskip
A combined experimental and computational study of the recrystallization process induced by swift heavy 
ions in pre-damaged silicon carbide crystals was reported in a recent paper by Debelle \emph{et al.} [Phys.\,Rev.\,B\,86,\,100102(R)\,(2012)]. 
In this study, the authors tried to mimic by means of molecular dynamic simulations both damage production induced by 
low energy ion irradiation in SiC and the recrystallization effect generated by subsequent swift heavy ion irradiation. 
Here we show that the simulations performed by the authors are far from being realistic and cannot reproduce, even qualitatively, the experimental results. In fact, in their simulation of damage production, amorphization is reached at an amount of deposited 
energy per target atom ($\sim$5.4 eV/atom) which is nearly \emph{5~times~less} than what is found in previous experimental and computational studies, 
whereas for the simulation of the recrystallization process, the recrystallization rate per incident ion is about \emph{40~times~higher }than the 
experimental value. Because of these extremely huge discrepancies, these molecular dynamic calculations are completely erroneous and the authors cannot claim that they found an ``exceptionally good 
agreement between experiments and simulations'' and that ``the recovery process is unambiguously accounted for by the thermal 
spike phenomenon''. In addition, no mention is given about a very similar experimental study recently published by some of the 
authors but with a quite different result for the case of fully amorphous SiC. 
\medskip
\begin{flushright}
PACS number(s): 61.80.-x, 61.43.Bn, 61.72.Cc, 61.85.+p \hspace*{1.8cm}
\medskip
\smallskip
\end{flushright}  
\end{abstract}    

\maketitle

The paper of Debelle  \emph{et al.} \cite{PhysRevB.86.100102} reports a combined experimental and computational study 
of the damage produced in silicon carbide (SiC) by ion irradiation with 100-keV Fe ions 
and its subsequent recrystallization by 870-MeV Pb ion irradiation. In order to reproduce 
the experimental results, the authors used molecular dynamic (MD) simulations for both 
damage production by the low energy ions and damage recovery induced by the thermal 
spike generated by the swift heavy ions. 

In an attempt to mimic the damage in SiC resulting from the nuclear collisions produced by 
the Fe irradiation at two different fluences ($2\times10^{14}$ and $4\times10^{14}$ Fe/cm$^2$), the authors 
generated a damaged layer by giving a kinetic energy $E$ = 50 eV to atoms (hereafter called recoils) 
randomly located within a Gaussian distribution with a standard deviation $\sigma$ = 2 nm centered in the middle (along the z-axis) of a 
cubic simulation box having a side length of 24 nm. The authors claimed that with this 
procedure they obtained a fully amorphous layer after 70000 recoils (see Fig.~2(a) of Ref.~1) and 
a damage peak with the maximum practically reaching total disorder (i.e., corresponding to 
the so-called ``amorphization threshold'' \footnote{In damage studies by ion irradiation, it is commonly admitted that the amorphization threshold is attained once the top of the disorder peak reaches total disorder.}) after 30000 recoils (see Fig.~2(b) of Ref.~1). These 
simulations are quite unrealistic. As a matter of fact, in a simulation with a total number of $N_{rec}$ recoils of kinetic energy $E$ randomly generated according to a Gaussian distribution with a standard deviation $\sigma$, the average energy deposited per target atom at the maximum of the distribution (neglecting the broadening caused by the recoils) is given by: 
\begin{equation}
\label{equation1}
\epsilon=\frac{1}{\sqrt{2\pi}\sigma}\frac{N_{rec}}{\Delta X \Delta Y}\frac{E}{\rho}
\end{equation}
where $\Delta X$ and $\Delta Y$ are the lateral dimensions of the simulation box and $\rho$ is the atomic density of the simulated material 
(9.64$\times10^{22}$ atom/cm$^3$ for SiC). The application of  Eq.~(\ref{equation1}) to the case of $N_{rec}$=30000 recoils indicates that in the simulations of Ref.~1 the threshold 
for amorphization is attained when the amount of deposited energy per target atom reaches about 5.4 eV/atom. 
This is more than \emph{5~times~less} than what is found experimentally in numerous previous studies which reported 
that the threshold for amorphization by low energy ion irradiation at room temperature is around 29 eV/atom.\cite {wendler1998ion} 
In addition, it must be emphasized that in a subsequent paper devoted only to the MD simulations \cite{backman2013molecular} a flat and 
completely amorphous layer was obtained after irradiation at room temperature with 24000 recoils of 50 eV homogeneously distributed in a 
4 nm-thick layer within a simulation box having a side length of 24 nm. It is thus evident that this complete amorphization 
is reached at $\frac{24000}{24\times24\times4\times10^{-21}}\frac{50}{9.64\times10^{22}}$ eV/atom = 5.4 eV/atom (neglecting here again 
the broadening caused by the recoils). 
This is again a striking evidence that the MD 
simulations performed by the authors have definitely an amorphization threshold corresponding to an amount of deposited 
energy per target atom which is more than 5 times less than the experimental value. This strong discrepancy becomes even 
more surprising if one realizes that about ten years ago Gao \emph{et al.} \cite {gao2002defect} were able to perform MD simulations describing the 
development of disorder induced in SiC by low energy ion irradiation where they found an amorphization threshold of 27.5 eV/atom 
quite close to the experimental values. \cite {wendler1998ion} It must also be emphasized that even for the simulation with $N_{rec}$=70000 recoils of Ref.~1, the formation of a fully amorphized layer cannot be achieved since in this case the energy deposited per atom at the center of the disordered layer (Eq.~(\ref{equation1})) is only 12.6 eV/atom, well below the amorphization threshold of $\sim$29 eV/atom. Therefore, owing to the very low energy densities 
of, respectively, 5.4 and 12.6 eV/atom used in the simulations supposed to reproduce the two experimental irradiations 
with $2\times10^{14}$ Fe/cm$^2$ for the partially (i.e., 98 \%) amorphous sample and with $4\times10^{14}$ Fe/cm$^2$ 
for the fully (i.e., 100 \%) amorphous sample, the disorder at the peak maximum is expected to be at only 10.3 \% for the first simulation and at only 23.3 \% for the second one, as shown in Fig.~1. The latter describes the evolution of 
the amount of disorder at the maximum of the damage peak with the deposited energy density deduced from the experimental irradiation of SiC with 100-keV Fe ions.\cite{thome2011radiation}
As a result, the two damage states of 98 \% and 100 \% of disorder in SiC can never be achieved in reality with the simulation conditions 
used in Ref.~1. Moreover, these serious misconceptions inevitably undermine the subsequent MD simulations of the swift heavy ion induced 
recrystallization effect (i.e., the objective of the paper) since the latter phenomenon is well known to be very sensitive to the actual initial 
damage level.\cite{benyagoub2009mechanism}

\begin{figure}
\includegraphics[scale=0.132, bb=0 0 1739 1364]{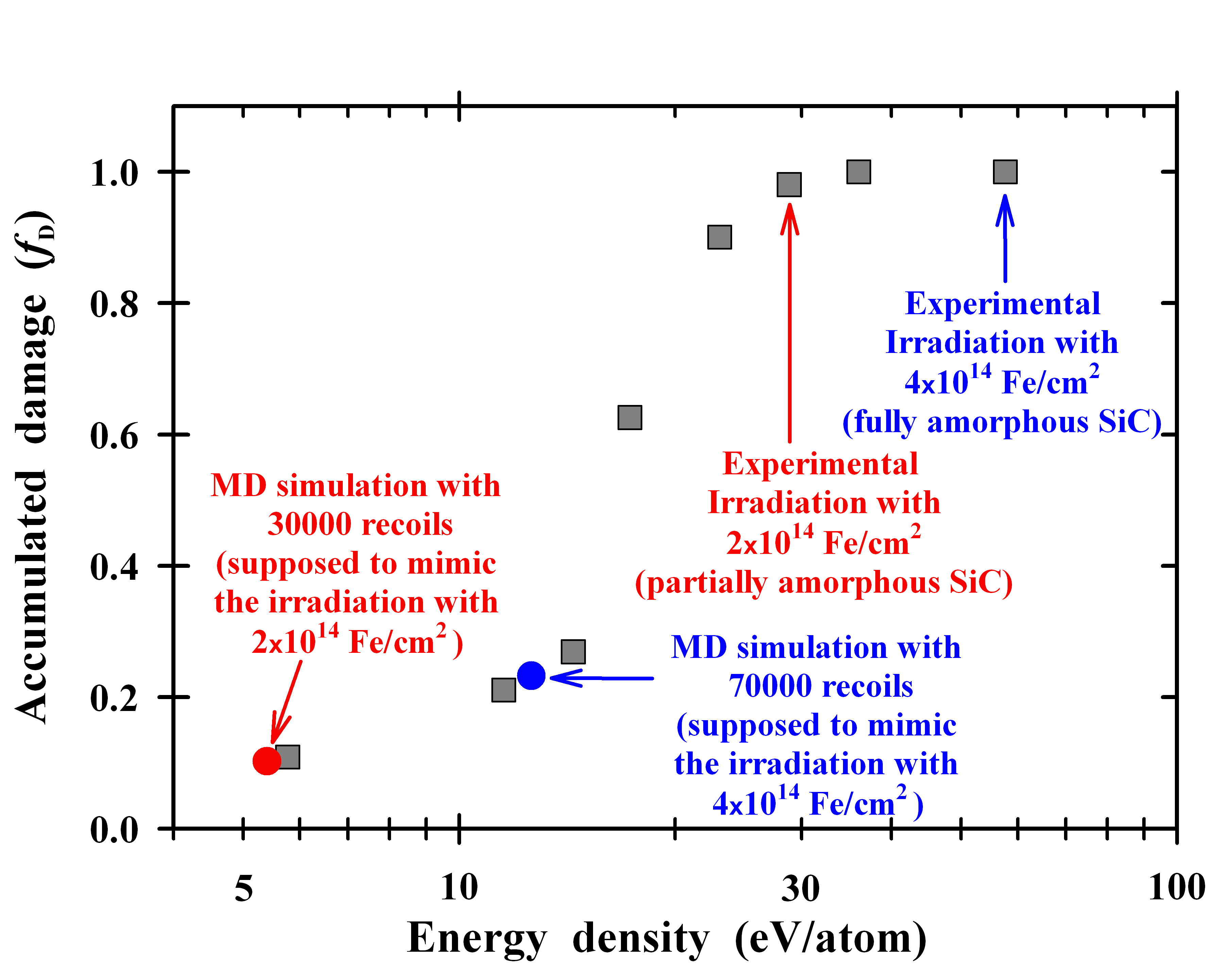}
\caption{Evolution of the experimental amount of disorder at the maximum of the damage peak with the deposited energy density produced by 100-keV Fe ion irradiation in SiC (experimental data taken from Ref.~6). The damage expected from the MD simulations with 30000 and 70000 recoils is clearly less than that induced by the experimental irradiations with $2\times10^{14}$ and $4\times10^{14}$ Fe/cm$^2$.}
\label{FIG. 1}
\end{figure}
\smallskip
In addition, it is well known that the disorder created by 50 eV Si or C recoils (i.e., just above the displacement threshold of 35 eV for Si and 21 eV for C) \cite{devanathan2000displacement} consists essentially in few isolated point defects and, as such, can hardly simulate the damage caused by 100-keV Fe ions where the Si and C primary knock-on atoms have median energies of the order of a few keV giving then rise to the generation of collision cascades containing hundreds of atoms. Actually, the authors' simulations are more representative of disorder creation by MeV electrons than by 100-keV Fe ions. Consequently, a more realistic MD simulation of the disorder build-up would have used higher energy recoils (of a few keV like in the MD calculations of Gao \emph{et al.} \cite{gao2002defect,gao2002cascade}) in order to be considered as relevant to low energy ion irradiation experiments at room temperature. In any case, the MD simulations of the amorphization process by 50 eV recoils performed in Ref. 1 are not in line with those previously reported by Malerba \emph{et al.} \cite{malerba2001molecular} using 100 eV recoils at 20~K in SiC. These authors found a critical energy density for amorphization of 23.4 eV/atom which is in agreement with experimental values determined at \emph{cryogenic temperatures} ($\sim$20 eV/atom). \cite{wendler1998ion}  Since according to numerous previous studies one naturally expects a higher amorphization threshold at room temperature than at cryogenic temperatures (see, e.g., Ref.~3), it is again quite surprising to find that the MD simulations of damage build-up by low energy recoils performed \emph{at room temperature} in Ref.~1 have an amorphization threshold (5.4 eV/atom) which is \emph{several times lower} than that (23.4 eV/atom) obtained at \emph{20~K} by Malerba \emph{et al.} \cite{malerba2001molecular}
\begin{figure*}
\includegraphics[scale=0.68, bb=0 0 698 281]{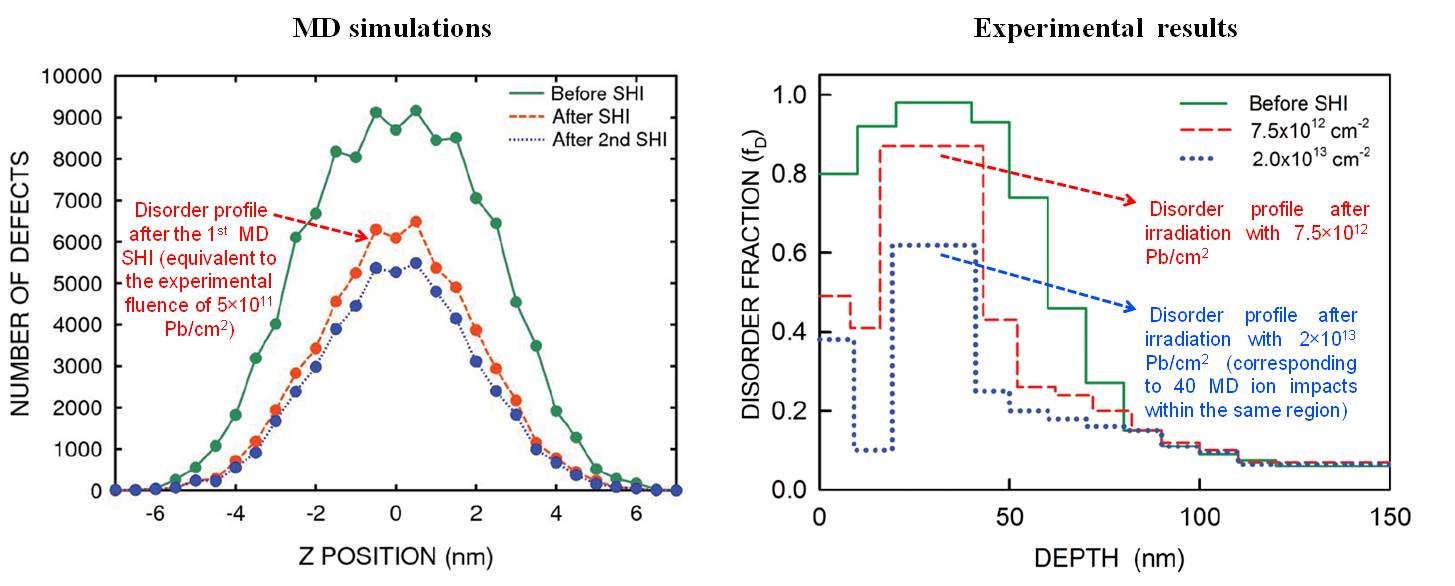}
\caption{Side-by-side comparison between Fig.~2(b) and Fig.~1(b) of Ref.~1 indicating similar recrystallization effect at the maximum of the disorder peak for the ``partially amorphous layer'' between the MD simulation of the first SHI impact (red curve on the left panel) and the experimental irradiation with $2\times10^{13}$ Pb/cm$^2$ (blue curve on the right panel). This simple comparison clearly shows that the recrystallization induced by the MD simulation of the first ion impact is about \emph{40~times~higher} than in the experiment.}
\label{FIG. 2}
\end{figure*}

In their MD simulation of the thermal spike caused by a 0.87-GeV Pb ion (giving rise to an electronic energy loss of $\sim$33 keV/nm), the authors added to the lattice atoms a cylindrical radial kinetic energy profile (deduced from the inelastic thermal spike model \cite{toulemonde2000transient}) and examined damage recovery within the most affected area, namely a central region having an effective swift heavy ion (SHI) track radius of 8 nm (the same radius was also used in Ref.~4). In their discussion, they \emph{directly} likened what happened in the MD cells after the first and second SHI impacts for both initial damage states (see Fig.~2 of Ref.~1) to what happened respectively after the experimental irradiations with $7.5\times10^{12}$ and $2\times10^{13}$ Pb/cm$^2$ (see Fig.~1 of Ref.~1) and concluded by claiming that they found an ``exceptionally good agreement between experiments and simulations''. This claim is actually erroneous. The authors \emph{overlooked} the fact that if each Pb ion induces damage recovery within a region having an effective track radius of 8 nm, this region will receive on average about 15 ion impacts at the fluence of $7.5\times10^{12}$ Pb/cm$^2$ instead of only one impact and about 40 impacts at the fluence of $2\times10^{13}$ Pb/cm$^2$ instead of only two ion impacts. \footnote{In the case of a homogeneously irradiated sample with an ion fluence $\Phi$, the average number $<n>$ of ion impacts within an area $\sigma$ normal to the beam direction is given by: $<n> = \sigma \Phi$.} In other words, each ion impact within this region corresponds in fact to an experimental fluence of $5\times10^{11}$ Pb/cm$^2$ and this conversion factor needs to be used in any comparison between the MD simulation and the experiment. Actually, a simple visual side-by-side comparison (see Fig.~2) between Fig.~2(b) and Fig.~1(b) of Ref.~1 reveals that, at the maximum of 
the damage peak, the first ion impact MD simulation induces a \emph{relative disorder decrease} significantly larger than that produced by the experimental irradiation with $7.5\times10^{12}$ Pb/cm$^2$ but quite close to that generated by the irradiation with $2\times10^{13}$ Pb/cm$^2$. With the above-mentioned conversion factor in mind, this simple comparison clearly shows ---without the need of any additional data--- that the recrystallization induced by the first ion impact simulation is approximately \emph{40~times~higher} than the experimental value. Therefore, the claim that the authors got an exceptionally good agreement between simulations and experiments is as wrong as to claim that the Moon and Jupiter have the same radius because they have the same size on two photos taken without reference to the scale. A more rigorous calculation can be performed by extracting from the experimental data the average recrystallization rate after each SHI irradiation fluence $\Phi_i$. This recrystallization rate, expressed in number of target atoms per incident ion per unit ion path length, is given by \cite{benyagoub2009mechanism}:
\begin{equation}
\label{equation2}
N(z,\Phi_i)=\frac{D(z,\Phi_{i-1})-D(z,\Phi_i)}{\Phi_i-\Phi_{i-1}}\rho
\end{equation}
where $D(z,\Phi_i)$ is the amount of disorder at depth $z$ and fluence $\Phi_i$ and $\rho$ is the SiC atomic density. The application of  Eq.~(\ref{equation2}) to the experimental data presented in Fig.~1(b) of Ref.~1 (see the right panel of Fig.~2) provides an average recrystallization rate at the maximum of the disorder peak of $\sim$140 atoms/nm per incident ion after the first Pb fluence. This number is in agreement with those found in a previous detailed study of the recrystallization effect induced by swift heavy (i.e., 827-MeV Pb) ions in pre-damaged SiC (see Fig.~4 of Ref.~7). In the case of the MD simulations, the recrystallization rate can also be deduced from the evolution of the simulated disorder profiles after the SHI irradiation. For instance, the MD recrystallization rate after the first ion impact for the ``partially amorphous layer'' is directly given by the difference between the green and the red curves in Fig.~2(b) of Ref.~1 (see the left panel of Fig.~2). By taking into account the fact that in this figure the depth interval is 0.5 nm, this gives a recrystallization rate of $\sim$5430 atoms/nm at the maximum of the disorder peak. This number is about 39 times higher than the experimental value ($\sim$140 atoms/nm), indicating clearly that the MD simulations of the thermal spike performed in Ref.~1 are definitely erroneous and are far from reproducing the experimental results.

Besides the huge discrepancy in the magnitude of the recrystallization process between the MD simulations and experiment, it can also be seen for the ``partially amorphous layer'' that at the maximum of the disorder peak the recovery process is more rapid after the first ion impact than the second one in the MD simulations in contrast to the experimental results shown in Fig.~1(b) of Ref.~1 and in Fig.~2(b) of Ref.~7 where several 827-MeV Pb fluences were used. These experimental data also reveal that damage recovery is initially more rapid at the wings of the disorder distribution than at the peak maximum (see Fig.~2(b) of Ref.~7 and also Fig.~1(b) of Ref.~1) contrary to the MD simulations where the disorder recovery occurs uniformly over the whole damaged layer. It is thus evident that the MD simulations reported in Ref.~1 are also not able to reproduce any of these very characteristic features of the SHI induced recrystallization phenomenon. 

In addition, although the greatest advantage of MD simulations is to provide information which is inaccessible to the experiment, these calculations do not give any detail on the intermediate structural modifications generated by the thermal spike. This is in contrast with previous thermal spike calculations \cite{benyagoub2008irradiation} which showed that 827-MeV Pb ions (giving rise to an electronic energy loss of $\sim$33 keV/nm) cannot amorphize crystalline SiC but can induce melting in amorphous SiC and the recrystallization process was explained \cite{benyagoub2009mechanism} by a mechanism combining the melting within the ion tracks of the amorphous zones through a thermal spike process and their subsequent epitaxial recrystallization initiated from the neighboring crystalline regions if the size of the latter surpasses a certain critical value.

Last but not least, it is surprising to find that in a recent paper \cite{thome2011radiation} a very similar experimental study to that described in Ref.~1 (see Figs.~11 and 12 of Ref.~6) was also published by some of the authors but with a quite different result. More precisely, it is reported in Ref.~6 that the fully amorphous SiC sample (obtained by the irradiation with 100-keV Fe ions at the fluence of $4\times10^{14}$ Fe/cm$^2$) exhibits a clear damage recovery, evidenced by the decrease of the damage fraction from 1 to $\sim$0.7, after subsequent irradiation at room temperature with 870-MeV Pb ions at the fluence of $2\times10^{13}$ Pb/cm$^2$ (see Fig.~12 of Ref.~6). This is plainly in contrast with the results shown in Fig.~1(a) of the authors' paper where one can see that there is no damage recovery at the disorder maximum for the fully amorphous sample. Consequently, two quite contradictory experimental results about the behavior of the fully amorphous SiC under swift heavy ion irradiation are now published in the literature by the same experimental group. This can only confuse the reader and may cast doubt on the reliability of their experimental results.
 
\medskip
In conclusion, the MD simulations performed by Debelle \emph{et al.} \cite{PhysRevB.86.100102} in order to mimic damage production by low energy ion irradiation as well as the recrystallization process induced by subsequent SHI irradiation are unrealistic. The authors clearly failed to reproduce either qualitatively or quantitatively their own experimental data. More precisely, there are, between the simulations and experiments, a discrepancy of more than 500 \% for the amorphization process and a huge discrepancy of about 4000 \% for the swift heavy ion induced recrystallization effect. The presence of these serious deficiencies undermines the authors' claim that they unambiguously explained the mechanism of the recrystallization phenomenon by MD simulations. Except the first publication in this research area, the authors ignored several results published in previous works and treated the subject as if it is a virgin field, avoiding thus any comparison of their data with those arising from the literature. Moreover, because as stated in the introduction of the authors' paper SiC is an important material for ``nuclear and space applications, in nuclear structural applications, and as an accident-tolerant cladding to prevent Fukushima-type accidents'', any simulation dealing with radiation-resistance of this material ought to be irreproachable due to the strong societal concern in this area. As a matter of fact, if the prediction of a \emph{single} phenomenon (such as the amorphization or the recrystallization process) can be so wrong, how can people trust on long-term simulations of nuclear waste repository sites (such that envisaged in Yucca Mountain (USA), Bure (France), Gorleben (Germany) and Honorobe (Japan)) where \emph{several} physical, chemical, geological and environmental processes are involved in a complex manner over several hundred thousand years? This may increase drastically the public suspicion against nuclear science and industry, especially after the accidents of Chernobyl and Fukushima.

\nocite{*}

\bibliography{SHI-Recryst-bib}

\begin{thebibliography}{13}%
\makeatletter
\providecommand \@ifxundefined [1]{%
 \@ifx{#1\undefined}
}%
\providecommand \@ifnum [1]{%
 \ifnum #1\expandafter \@firstoftwo
 \else \expandafter \@secondoftwo
 \fi
}%
\providecommand \@ifx [1]{%
 \ifx #1\expandafter \@firstoftwo
 \else \expandafter \@secondoftwo
 \fi
}%
\providecommand \natexlab [1]{#1}%
\providecommand \enquote  [1]{``#1''}%
\providecommand \bibnamefont  [1]{#1}%
\providecommand \bibfnamefont [1]{#1}%
\providecommand \citenamefont [1]{#1}%
\providecommand \href@noop [0]{\@secondoftwo}%
\providecommand \href [0]{\begingroup \@sanitize@url \@href}%
\providecommand \@href[1]{\@@startlink{#1}\@@href}%
\providecommand \@@href[1]{\endgroup#1\@@endlink}%
\providecommand \@sanitize@url [0]{\catcode `\\12\catcode `\$12\catcode
  `\&12\catcode `\#12\catcode `\^12\catcode `\_12\catcode `\%12\relax}%
\providecommand \@@startlink[1]{}%
\providecommand \@@endlink[0]{}%
\providecommand \url  [0]{\begingroup\@sanitize@url \@url }%
\providecommand \@url [1]{\endgroup\@href {#1}{\urlprefix }}%
\providecommand \urlprefix  [0]{URL }%
\providecommand \Eprint [0]{\href }%
\providecommand \doibase [0]{http://dx.doi.org/}%
\providecommand \selectlanguage [0]{\@gobble}%
\providecommand \bibinfo  [0]{\@secondoftwo}%
\providecommand \bibfield  [0]{\@secondoftwo}%
\providecommand \translation [1]{[#1]}%
\providecommand \BibitemOpen [0]{}%
\providecommand \bibitemStop [0]{}%
\providecommand \bibitemNoStop [0]{.\EOS\space}%
\providecommand \EOS [0]{\spacefactor3000\relax}%
\providecommand \BibitemShut  [1]{\csname bibitem#1\endcsname}%
\let\auto@bib@innerbib\@empty
\bibitem [{\citenamefont {Debelle}\ \emph {et~al.}(2012)\citenamefont
  {Debelle}, \citenamefont {Backman}, \citenamefont {Thom\'e}, \citenamefont
  {Weber}, \citenamefont {Toulemonde}, \citenamefont {Mylonas}, \citenamefont
  {Boulle}, \citenamefont {Pakarinen}, \citenamefont {Juslin}, \citenamefont
  {Djurabekova}, \citenamefont {Nordlund}, \citenamefont {Garrido},\ and\
  \citenamefont {Chaussende}}]{PhysRevB.86.100102}%
  \BibitemOpen
  \bibfield  {author} {\bibinfo {author} {\bibfnamefont {A.}~\bibnamefont
  {Debelle}}, \bibinfo {author} {\bibfnamefont {M.}~\bibnamefont {Backman}},
  \bibinfo {author} {\bibfnamefont {L.}~\bibnamefont {Thom\'e}}, \bibinfo
  {author} {\bibfnamefont {W.~J.}\ \bibnamefont {Weber}}, \bibinfo {author}
  {\bibfnamefont {M.}~\bibnamefont {Toulemonde}}, \bibinfo {author}
  {\bibfnamefont {S.}~\bibnamefont {Mylonas}}, \bibinfo {author} {\bibfnamefont
  {A.}~\bibnamefont {Boulle}}, \bibinfo {author} {\bibfnamefont {O.~H.}\
  \bibnamefont {Pakarinen}}, \bibinfo {author} {\bibfnamefont {N.}~\bibnamefont
  {Juslin}}, \bibinfo {author} {\bibfnamefont {F.}~\bibnamefont {Djurabekova}},
  \bibinfo {author} {\bibfnamefont {K.}~\bibnamefont {Nordlund}}, \bibinfo
  {author} {\bibfnamefont {F.}~\bibnamefont {Garrido}}, \ and\ \bibinfo
  {author} {\bibfnamefont {D.}~\bibnamefont {Chaussende}},\ }\href {\doibase
  10.1103/PhysRevB.86.100102} {\bibfield  {journal} {\bibinfo  {journal} {Phys.
  Rev. B}\ }\textbf {\bibinfo {volume} {86}},\ \bibinfo {pages} {100102(R)}
  (\bibinfo {year} {2012})}\BibitemShut {NoStop}%
\bibitem [{Note1()}]{Note1}%
  \BibitemOpen
  \bibinfo {note} {In damage studies by ion irradiation, it is commonly
  admitted that the amorphization threshold is attained once the top of the
  disorder peak reaches total disorder.}\BibitemShut {Stop}%
\bibitem [{\citenamefont {Wendler}\ \emph {et~al.}(1998)\citenamefont
  {Wendler}, \citenamefont {Heft},\ and\ \citenamefont
  {Wesch}}]{wendler1998ion}%
  \BibitemOpen
  \bibfield  {author} {\bibinfo {author} {\bibfnamefont {E.}~\bibnamefont
  {Wendler}}, \bibinfo {author} {\bibfnamefont {A.}~\bibnamefont {Heft}}, \
  and\ \bibinfo {author} {\bibfnamefont {W.}~\bibnamefont {Wesch}},\
  }\href@noop {} {\bibfield  {journal} {\bibinfo  {journal} {Nuclear
  Instruments and Methods in Physics Research Section B: Beam Interactions with
  Materials and Atoms}\ }\textbf {\bibinfo {volume} {141}},\ \bibinfo {pages}
  {105} (\bibinfo {year} {1998})}\BibitemShut {NoStop}%
\bibitem [{\citenamefont {Backman}\ \emph {et~al.}(2013)\citenamefont
  {Backman}, \citenamefont {Toulemonde}, \citenamefont {Pakarinen},
  \citenamefont {Juslin}, \citenamefont {Djurabekova}, \citenamefont
  {Nordlund}, \citenamefont {Debelle}, \citenamefont {Weber} \emph
  {et~al.}}]{backman2013molecular}%
  \BibitemOpen
  \bibfield  {author} {\bibinfo {author} {\bibfnamefont {M.}~\bibnamefont
  {Backman}}, \bibinfo {author} {\bibfnamefont {M.}~\bibnamefont {Toulemonde}},
  \bibinfo {author} {\bibfnamefont {O.~H.}\ \bibnamefont {Pakarinen}}, \bibinfo
  {author} {\bibfnamefont {N.}~\bibnamefont {Juslin}}, \bibinfo {author}
  {\bibfnamefont {F.}~\bibnamefont {Djurabekova}}, \bibinfo {author}
  {\bibfnamefont {K.}~\bibnamefont {Nordlund}}, \bibinfo {author}
  {\bibfnamefont {A.}~\bibnamefont {Debelle}}, \bibinfo {author} {\bibfnamefont
  {W.~J.}\ \bibnamefont {Weber}},  \emph {et~al.},\ }\href@noop {} {\bibfield
  {journal} {\bibinfo  {journal} {Computational Materials Science}\ }\textbf
  {\bibinfo {volume} {67}},\ \bibinfo {pages} {261} (\bibinfo {year}
  {2013})}\BibitemShut {NoStop}%
\bibitem [{\citenamefont {Gao}\ \emph {et~al.}(2002)\citenamefont {Gao},
  \citenamefont {Weber},\ and\ \citenamefont {Devanathan}}]{gao2002defect}%
  \BibitemOpen
  \bibfield  {author} {\bibinfo {author} {\bibfnamefont {F.}~\bibnamefont
  {Gao}}, \bibinfo {author} {\bibfnamefont {W.~J.}\ \bibnamefont {Weber}}, \
  and\ \bibinfo {author} {\bibfnamefont {R.}~\bibnamefont {Devanathan}},\
  }\href@noop {} {\bibfield  {journal} {\bibinfo  {journal} {Nuclear
  Instruments and Methods in Physics Research Section B: Beam Interactions with
  Materials and Atoms}\ }\textbf {\bibinfo {volume} {191}},\ \bibinfo {pages}
  {487} (\bibinfo {year} {2002})}\BibitemShut {NoStop}%
\bibitem [{\citenamefont {Thom{\'e}}\ \emph {et~al.}(2012)\citenamefont
  {Thom{\'e}}, \citenamefont {Moll}, \citenamefont {Debelle}, \citenamefont
  {Garrido}, \citenamefont {Sattonnay},\ and\ \citenamefont
  {Jagielski}}]{thome2011radiation}%
  \BibitemOpen
  \bibfield  {author} {\bibinfo {author} {\bibfnamefont {L.}~\bibnamefont
  {Thom{\'e}}}, \bibinfo {author} {\bibfnamefont {S.}~\bibnamefont {Moll}},
  \bibinfo {author} {\bibfnamefont {A.}~\bibnamefont {Debelle}}, \bibinfo
  {author} {\bibfnamefont {F.}~\bibnamefont {Garrido}}, \bibinfo {author}
  {\bibfnamefont {G.}~\bibnamefont {Sattonnay}}, \ and\ \bibinfo {author}
  {\bibfnamefont {J.}~\bibnamefont {Jagielski}},\ }\href@noop {} {\bibfield
  {journal} {\bibinfo  {journal} {Advances in Materials Science and
  Engineering}\ }\textbf {\bibinfo {volume} {2012}},\ \bibinfo {pages} {905474}
  (\bibinfo {year} {2012})}\BibitemShut {NoStop}%
\bibitem [{\citenamefont {Benyagoub}\ and\ \citenamefont
  {Audren}(2009)}]{benyagoub2009mechanism}%
  \BibitemOpen
  \bibfield  {author} {\bibinfo {author} {\bibfnamefont {A.}~\bibnamefont
  {Benyagoub}}\ and\ \bibinfo {author} {\bibfnamefont {A.}~\bibnamefont
  {Audren}},\ }\href@noop {} {\bibfield  {journal} {\bibinfo  {journal}
  {Journal of Applied Physics}\ }\textbf {\bibinfo {volume} {106}},\ \bibinfo
  {pages} {083516} (\bibinfo {year} {2009})}\BibitemShut {NoStop}%
\bibitem [{\citenamefont {Devanathan}\ and\ \citenamefont
  {Weber}(2000)}]{devanathan2000displacement}%
  \BibitemOpen
  \bibfield  {author} {\bibinfo {author} {\bibfnamefont {R.}~\bibnamefont
  {Devanathan}}\ and\ \bibinfo {author} {\bibfnamefont {W.~J.}\ \bibnamefont
  {Weber}},\ }\href@noop {} {\bibfield  {journal} {\bibinfo  {journal} {Journal
  of nuclear materials}\ }\textbf {\bibinfo {volume} {278}},\ \bibinfo {pages}
  {258} (\bibinfo {year} {2000})}\BibitemShut {NoStop}%
\bibitem [{\citenamefont {Gao}\ and\ \citenamefont
  {Weber}(2002)}]{gao2002cascade}%
  \BibitemOpen
  \bibfield  {author} {\bibinfo {author} {\bibfnamefont {F.}~\bibnamefont
  {Gao}}\ and\ \bibinfo {author} {\bibfnamefont {W.~J.}\ \bibnamefont
  {Weber}},\ }\href@noop {} {\bibfield  {journal} {\bibinfo  {journal}
  {Physical Review B}\ }\textbf {\bibinfo {volume} {66}},\ \bibinfo {pages}
  {024106} (\bibinfo {year} {2002})}\BibitemShut {NoStop}%
\bibitem [{\citenamefont {Malerba}\ and\ \citenamefont
  {Perlado}(2001)}]{malerba2001molecular}%
  \BibitemOpen
  \bibfield  {author} {\bibinfo {author} {\bibfnamefont {L.}~\bibnamefont
  {Malerba}}\ and\ \bibinfo {author} {\bibfnamefont {J.}~\bibnamefont
  {Perlado}},\ }\href@noop {} {\bibfield  {journal} {\bibinfo  {journal}
  {Journal of nuclear materials}\ }\textbf {\bibinfo {volume} {289}},\ \bibinfo
  {pages} {57} (\bibinfo {year} {2001})}\BibitemShut {NoStop}%
\bibitem [{\citenamefont {Toulemonde}\ \emph {et~al.}(2000)\citenamefont
  {Toulemonde}, \citenamefont {Dufour}, \citenamefont {Meftah},\ and\
  \citenamefont {Paumier}}]{toulemonde2000transient}%
  \BibitemOpen
  \bibfield  {author} {\bibinfo {author} {\bibfnamefont {M.}~\bibnamefont
  {Toulemonde}}, \bibinfo {author} {\bibfnamefont {C.}~\bibnamefont {Dufour}},
  \bibinfo {author} {\bibfnamefont {A.}~\bibnamefont {Meftah}}, \ and\ \bibinfo
  {author} {\bibfnamefont {E.}~\bibnamefont {Paumier}},\ }\href@noop {}
  {\bibfield  {journal} {\bibinfo  {journal} {Nuclear Instruments and Methods
  in Physics Research Section B: Beam Interactions with Materials and Atoms}\
  }\textbf {\bibinfo {volume} {166}},\ \bibinfo {pages} {903} (\bibinfo {year}
  {2000})}\BibitemShut {NoStop}%
\bibitem [{Note2()}]{Note2}%
  \BibitemOpen
  \bibinfo {note} {In the case of a homogeneously irradiated sample with an ion
  fluence $\Phi $, the average number $<n>$ of ion impacts within an area
  $\sigma $ normal to the beam direction is given by: $<n> = \sigma \Phi
  $.}\BibitemShut {Stop}%
\bibitem [{\citenamefont {Benyagoub}(2008)}]{benyagoub2008irradiation}%
  \BibitemOpen
  \bibfield  {author} {\bibinfo {author} {\bibfnamefont {A.}~\bibnamefont
  {Benyagoub}},\ }\href@noop {} {\bibfield  {journal} {\bibinfo  {journal}
  {Nuclear Instruments and Methods in Physics Research Section B: Beam
  Interactions with Materials and Atoms}\ }\textbf {\bibinfo {volume} {266}},\
  \bibinfo {pages} {2766} (\bibinfo {year} {2008})}\BibitemShut {NoStop}%
\end{thebibliography}%

\end{document}